\documentstyle[12pt,epsfig]{article}

\newcommand{\beeq}{\begin{equation}}
\newcommand{\bear}{\begin{eqnarray}}
\newcommand{\eneq}{\end{equation}}
\newcommand{\enar}{\end{eqnarray}}

\newcommand{\gsim}{\;\raisebox{-0.9ex}
           {$\textstyle\stackrel{\textstyle >}{\sim}$}\;}

\newcommand{\lsim}{\;\raisebox{-0.9ex}{$\textstyle\stackrel{\textstyle<}
           {\sim}$}\;}

\newcommand{\non}{\nonumber}

\newcommand{\msi}{m_{{\tilde{q}_i}}}
\newcommand{\msj}{m_{{\tilde{q}_j}}}
\newcommand{\msg}{m_{\tilde{g}}}
\newcommand{\Li}{\mbox{{\rm Li}$_2$}}

\newcommand{\heta}{{\theta_{\tilde q}}}

\def\sq{\ifmmode{\tilde{q}} \else{$\tilde{q}$} \fi}
\def\st{\ifmmode{\tilde{t}} \else{$\tilde{t}$} \fi}
\def\sb{\ifmmode{\tilde{b}} \else{$\tilde{b}$} \fi}
\def\sg{\ifmmode{\tilde{g}} \else{$\tilde{g}$} \fi}
\begin{document}

\pagestyle{empty}
%\vspace*{-3cm}
\begin{flushright}
UWThPh-1996-6\\
HEPHY-PUB 640/96\\
hep-ph/9603206\\
\vspace{0.3cm}
February, 1996
\end{flushright}

\vspace{1cm}
\begin{center}
\begin{Large} \bf
SUSY--QCD corrections to\\
scalar quark pair production\\
in \boldmath{$e^+ e^-$} annihilation\\
\end{Large}
\end{center}
\vspace{15mm}

\begin{center}
\large H.~Eberl$^1$, A.~Bartl$^2$, W.~Majerotto$^1$
\end{center}
\vspace{0mm}

\begin{center}
\begin{tabular}{l}
$^1${\it Institut f\"ur Hochenergiephysik der \"Osterreichischen Akademie
der Wissenschaften}\\
$^2${\it Institut f\"ur Theoretische Physik, Universit\"at Wien, A-1090
Vienna, Austria} 
\end{tabular}
\end{center}
%\vspace{10mm}
\vfill

\begin{abstract}
\begin{small}
\baselineskip=28pt
We calculate the supersymmetric ${\cal O}(\alpha_s)$ QCD corrections to the
cross section $e^+ e^- \to \sq_i \bar\sq_j$ $(i,j = 1, 2)$ within the
Minimal Supersymmetric Standard Model. We pay particular attention to the 
case of the left--right squark mixing and to the renormalization
of the mixing angle. The corrections due to gluino exchange turn out to be
smaller than those
due to gluon exchange, but they can be significant at higher energies
even for a gluino
mass of a few hundred GeV.
\end{small}
\end{abstract}

%\vfill

\newpage
\pagestyle{plain}
\setcounter{page}{2}
\baselineskip=21pt

\section{Introduction}
Supersymmetry \cite{1} requires the existence of two scalar partners
$\sq_L$ and $\sq_R$ (squarks) for every quark $q$. Quite generally, $\sq_L$ and
$\sq_R$ mix, the size of 
mixing being proportional to the mass of the quark \cite{2}. Therefore,
the scalar partners of the top quark are expected to be strongly mixed so that
one mass eigenstate $\st_1$ can be rather light. Within the Minimal 
Supersymmetric Standard Model (MSSM) $\sb_L$ and $\sb_R$ can
considerably mix \cite{3,3a} for large $\tan\beta = v_2/v_1$ (where $v_1$ 
and $v_2$ are the vacuum expectation values of the neutral components of the 
two Higgs doublets). Therefore, it is possible that $\st_1$ or $\sb_1$ will be
discovered in the energy range
of the present colliders.\\
The stop production in $e^+ e^-$ annihilation,  $e^+ e^- \to \st_1 \bar{\st}_1$,
was first studied at tree level in \cite{4}. The conventional QCD corrections to
this process including virtual and real (soft and hard) gluon radiation were
given in \cite{5,5a}. These corrections are quite large, for instance, they are
about 35\% for $m_{\st_1} = 80~$~GeV at $\sqrt{s} = 190$~GeV~\cite{6}. The QCD
corrections within the MSSM including virtual gluino and squark exchange were
first calculated in \cite{7}, where
the process $e^+ e^- \to \sq_1 \bar{\sq}_2$ was also included. However, the 
renormalization of the squark mixing angle $\heta$ 
as adopted there is not applicable in the
whole range of $\heta$ for the diagonal channels $e^+ e^- \to \sq_1 \bar{\sq}_1,
\,\sq_2 \bar{\sq}_2$.    
A proper renormalization of the squark mixing
angle is necessary whenever stop and sbottom
play a r\^ole. Moreover, no numerical analysis at all
has been given so far for the
unequal mass case or for squark mixing angle $\heta\neq 0$.\\

Here we want to present the complete formulae for the supersymmetric QCD 
corrections up to ${\cal O}(\alpha_s)$ within the MSSM including virtual and
real gluon exchange, virtual gluinos and squarks for general 
$\sq_L$--$\sq_R$ mixing. In
particular, we propose a suitable 
renormalization of the squark mixing angle.
Furthermore, we give a detailed
numerical analysis of these corrections.

\section{Tree level formulae}

The squark mixing of $\sq_L$ and $\sq_R$ is expressed by
\begin{eqnarray}
\tilde{q}_1=\tilde{q}_L \cos \heta +\tilde{q}_R \sin \heta \ \ , \hspace*{1cm}
\tilde{q}_2=- \tilde{q}_L \sin \heta +\tilde{q}_R \cos \heta \, ,
\end{eqnarray}
where  $\sq_1$,  $\sq_2$ are the mass eigenstates (with $m_{\sq_1} < m_{\sq_2}$)
and $\heta$ is the squark mixing angle. The production $e^+ e^- \to \sq_i
\bar{\sq}_j$, $(i,j = 1,2)$, proceeds via 
$\gamma$ and $Z$ exchange in the s--channel
(see Fig.~1a), $s =
(k + \bar{k})^2$, $k$ and $\bar{k}$ being the four--momenta of the outgoing
$\sq_i$ and $\bar{\sq}_j$.\\
The cross section at tree--level is given by:
\begin{equation} \label{sigtree}
\sigma^0(e^+ e^- \to \sq_i \bar{\sq}_j)
= c_{ij}\left[ e_q^2 \delta_{ij} - T_{\gamma Z} e_q a_{ij} \delta_{ij}
+ T_{ZZ} a_{ij}^2 \right] \, ,
\end{equation}
with
\begin{eqnarray}
c_{ij} & = & \frac{\pi \alpha^2}{s}\lambda^{3/2}_{ij} \, ,\label{cij}\\
T_{\gamma Z} & = &  \frac{v_e}
{8c_W^2s_W^2}\frac{s(s-m_Z^2)}{(s-m_Z^2)^2+ \Gamma^2_Z m_Z^2}\, ,
\label{TgZ}\\
T_{ZZ} & = & \frac{(a_e^2+v_e^2)}{256s_W^4c_W^4}
\frac{s^2}{(s-m_Z^2)^2+ \Gamma^2_Z m_Z^2} \label{TZZ}\, .
\end{eqnarray}
Here $\lambda_{ij}= (1-\mu_i^2 - \mu_j^2)^2 - 4 \mu_i^2 \mu_j^2,
\mu^2_{i,j}= m^2_{\sq_{i,j}}/s$; $e_q$ is the charge of the squarks
($e_t = 2/3, e_b = - 1/3$) in units of $e ( = \sqrt{4 \pi \alpha})$. $a_e$
and $v_e$ are the vector and axial vector couplings of the electron to the
$Z$ boson: $v_e = -1 + 4 s_W^2$ (with $s_W \equiv \sin\theta_W$), $a_e = -1$, 
and $a_{ij}$ are the corresponding couplings $Z \sq_i \bar{\sq}_j$:
\begin{eqnarray} \label{acoup}
a_{11}= 4(I_q^{3L} \cos^2 \heta -s_W^2 e_q) \ , \
a_{22}= 4(I_q^{3L} \sin^2 \heta -s_W^2 e_q) \ , \
a_{12}= a_{21}= -2 I_q^{3L} \sin 2\heta \, ,
\end{eqnarray}
where $I_q^{3L}$ is the third component of the weak isospin of the quark $q$.

\section{SUSY--QCD corrections}

The supersymmetric QCD corrected cross section in ${\cal O}(\alpha_s)$ 
corresponding to the Fig.~1 can be written as:
\begin{equation} \label{sig1}
\sigma = \sigma^0 + \delta \sigma^{g} + \delta \sigma^{\tilde g}
+ \delta \sigma^{\tilde q} \, .
\end{equation} 
$\delta \sigma^{g}$ gives the standard QCD gluonic correction
(Figs.~1b--f), $\delta \sigma^{\sg}$ is the correction 
due to the gluino exchange
(Figs.~1g, 1h) and
$\delta \sigma^{\sq}$ due to the squark exchange (Figs.~1i, 1j).
As will be seen later, within the renormalization prescription used
$\delta \sigma^{\sq} = 0$.\\ 
According to eq.~(\ref{sigtree}) the correction can be written as:
\begin{equation} \label{dsiga} \label{2.16}
\delta \sigma^{a} = c_{ij}\left[ 2 e_q \Delta (e_q)_{ij}^{(a)} \delta_{ij} -
 T_{\gamma Z} (e_q \delta_{ij}\Delta a_{ij}^{(a)} +
\Delta (e_q)_{ij}^{(a)} a_{ij})
+ 2 T_{ZZ} a_{ij} \Delta a_{ij}^{(a)} \right]
\quad ,\, a = g, \tilde g, \tilde q\, .
\eneq

\subsection{Gluonic correction}
$\delta \sigma^{g}$ can be written as
\begin{equation} \label{dsigg}
\delta \sigma^{g} = \sigma^0\left[\frac{4}{3}
\frac{\alpha_s}{\pi} \Delta_{ij}\right]\, .
\end{equation}
$\Delta_{ij}$ has the following expression:
\begin{eqnarray}\label{delgluon}
\Delta_{ij} &=& \log (\mu_i \mu_j) +2
+ \frac{2+\mu_i^2+\mu_j^2}{\lambda_{ij}^{1/2}} \log \lambda_0
+ \frac{1+2 \mu_i^2}{\lambda_{ij}^{1/2}}  \log \lambda_1 +
\frac{1+2\mu_j^2}{\lambda_{ij}^{1/2}} \log \lambda_2 \non\\
&+&
\frac{1- \mu_i^2-\mu_j^2}{\lambda_{ij}^{1/2}}
 \log \frac{1-\mu_i^2 -\mu_j^2+\lambda_{ij}^{1/2}}
{1-\mu_i^2 -\mu_j^2 -\lambda_{ij}^{1/2}} +
\left[\frac{(1-\mu_i^2-\mu_j^2)}{\lambda_{ij}^{1/2}} \log\lambda_0 -1
\right] \log \frac{\lambda_{ij}^2}{\mu_i^2 \mu_j^2} \\
&+&  \frac{4}{\lambda_{ij}
^{3/2}} \left[ \frac{1}{4} \lambda_{ij}^{1/2} (1+\mu_i^2+ \mu_j^2) + \mu_i^2
\log \lambda_2+ \mu_j^2 \log \lambda_1+ \mu_i^2 \mu_j^2 \log \lambda_0 \right]
+\frac{1-\mu_i^2-\mu_j^2}{\lambda_{ij}^{1/2}}\left[\frac{2\pi^2}{3}\right.\non\\
&+&\left. 2\Li(1-\lambda_0^2)+\Li(\lambda_1^2)- \Li(1-\lambda_1^2)+
\Li(\lambda_2^2)- \Li(1-\lambda_2^2) + 2 \log^2 \lambda_0 -\log\lambda_{ij}
\log\lambda_0\right]\non \, ,
\end{eqnarray}
with
\begin{eqnarray} \label{lam}
\lambda_0= \frac{1}{2\mu_i \mu_j}(1-\mu_i^2 - \mu_j^2 +\lambda_{ij}^{1/2})
\ \ \ \ , \ \
\lambda_{1,2}= \frac{1}{2\mu_{j,i}} (1 \mp \mu_i^2 \pm \mu_j^2 -\lambda_{ij}^{
1/2})\, ,
\end{eqnarray}
and $\Li(x) = - \int_0^1 \log (1-xt)/ t\, \rm{d}t$.
Eq.~(\ref{delgluon}) has been calculated from the graphs in Figs.~1b--f
including soft
and hard gluon radiation. For $i=j$ the expression eq.~(\ref{delgluon}) was
already derived in \cite{5,5a,7,schwinger}. 
For $i \neq j$ we also agree with the 
calculation of ref.~\cite{7} apart from an obvious misprint in eq.~(3.2) of
\cite{7}.

%\clearpage
\subsection{The four--squark interaction}
Quite generally, we can decompose the corrections due to the four--squark
interaction (Figs.~ 1i, 1j) in the following way:
\bear \label{deltaasq}
\Delta a_{ij}^{(\sq)} & = & \delta a_{ij}^{(v,\sq)} + \delta a_{ij}^{(w,\sq)} +
\delta a_{ij}^{(\tilde\theta,\sq)}\, ,\\
\Delta (e_q)_{ij}^{(\sq)} & = & 
\delta (e_q)_{ij}^{(v,\sq)} + \delta (e_q)_{ij}^{(w,\sq)}\label{deltaesq}\, .
\enar
Here the upper index $v$ denotes the vertex correction (Fig.~1i), 
$w$ the wave--function correction (Fig.~1j),
and $\tilde\theta$ the renormalization of the mixing angle $\heta$.
The last term in eq.~(\ref{deltaasq}) 
is necessary as the couplings $a_{ij}$ depend on the mixing angle (see 
eq.~(\ref{acoup})).
The vertex correction is proportional to the
momentum of the exchanged photon or $Z$ boson, and hence the related
matrix element for $e^+ e^- \to \sq_i \bar{\sq}_j$
is zero, that is $\delta
a_{ij}^{(v,\sq)} =
\delta (e_q)_{ij}^{(v,\sq)} = 0$. There is, however, a squark wave--function
correction $\delta a_{ij}^{(w,\sq)}$ due to the four--squark interaction.\\
The self--energies due to the squark loops in Fig.~1j are given by:
\beeq
\Sigma_{ij}^{(\tilde q)}(p^2) = \frac{1}{3} \frac{\alpha_s}{\pi} \left(
\begin{array}{cc}
\cos^2 2 \heta A^0(m_{{\tilde q}_1}^2) +
 \sin^2 2 \heta A^0(m_{{\tilde q}_2}^2)&
 \frac{1}{2}\sin 4 \heta
 \left(A^0(m_{{\tilde q}_2}^2)- A^0(m_{{\tilde q}_1}^2)\right)\\
 \frac{1}{2}\sin 4 \heta
 \left(A^0(m_{{\tilde q}_2}^2)- A^0(m_{{\tilde q}_1}^2)\right) &
\sin^2 2 \heta A^0(m_{{\tilde q}_1}^2)
 + \cos^2 2 \heta A^0(m_{{\tilde q}_2}^2)
\end{array}\right)\, .
\eneq
Here $A^0$ is the standard one--point function \cite{PaVe}, $A^0(m^2) =
-i \pi^{-2}\int d^Dq (q^2-m^2)^{-1}$ in the convention of \cite{8}.
Notice that the self--energies are independent of $p^2$, hence 
$\Sigma_{ij}^{(\tilde q)}(m_{{\tilde q}_1}^2)=
\Sigma_{ij}^{(\tilde q)}(m_{{\tilde q}_2}^2)=
\Sigma_{ij}^{(\tilde q)}$. We therefore get for 
$\delta a_{ij}^{(w,\sq)}$:
\beeq \label{15}
\delta a_{ij}^{(w,\tilde q)} =
\delta Z_{i'i}^{(\tilde q)} a_{i'j}
+\delta Z_{j'j}^{(\tilde q)} a_{ij'} =
-\mbox{Re}\left\{
\frac{\Sigma_{i'i}^{(\tilde q)}}
{m_{\sq_i}^2-m_{\sq_{i'}}^2} a_{i'j}
+ \frac{\Sigma_{j'j}^{(\tilde q)}
}{m_{\sq_j}^2-m_{\sq_{j'}}^2} a_{ij'}
\right\} \, ,\begin{array}{c} i \neq i' \\ j \neq j'\end{array}\, ,
\eneq
with the squark wave--function renormalization constants 
\begin{equation}
\delta Z_{12}^{(\sq)}=
\frac{\mbox{Re}\left\{\Sigma_{12}^{(\sq)}\right\}
}{m_{\tilde q_1}^2-
m_{\tilde q_2}^2} \quad , \quad
\delta Z_{21}^{(\sq)}=
\frac{\mbox{Re}\left\{\Sigma_{21}^{(\sq)}\right\}
}{m_{\tilde q_2}^2-
m_{\tilde q_1}^2} \, .
\end{equation}
Note that $\delta (e_q)_{ij}^{(w,\sq)} = 0$ because the contributions coming 
from the squark loop Fig.~1j and the corresponding graph with the loop at the
antisquark cancel each other. 
Hence $\Delta (e_q)_{ij}^{(\sq)} = 0$. The correction term
$\delta a_{ij}^{(\tilde\theta,\sq)}$ in eq.~(\ref{deltaasq})
will be treated in section 3.4. 

\subsection{Correction due to gluino exchange}
Now we turn to the corrections due to gluino exchange (Fig.~1g, 1h).
As in eq.~(\ref{deltaasq}) and eq.~(\ref{deltaesq}) we can write:
\bear \label{deltaasg}
\Delta a_{ij}^{(\sg)} &= & \delta a_{ij}^{(v,\sg)} + \delta a_{ij}^{(w,\sg)} +
\delta a_{ij}^{(\tilde\theta,\sg)}\, ,\\ \label{deltaesg}
\Delta (e_q)_{ij}^{(\sg)} &= & \delta (e_q)_{ij}^{(v,\sg)} + 
\delta (e_q)_{ij}^{(w,\sg)}\, ,
\enar 
with $\delta a_{ij}^{(v,\sg)},  
\delta (e_q)_{ij}^{(v,\sg)}$ corresponding to the
vertex correction (Fig.~1g), and
$\delta a_{ij}^{(w,\sg)}, \delta (e_q)_{ij}^{(w,\sg)}$
corresponding to the wave--function correction (Fig.~1h).
Again $\delta a_{ij}^{(\tilde\theta,\sg)}$ is due to the renormalization
of the mixing angle and will be calculated in section 3.4.\\
The gluino vertex corrections are given by:
\bear \label{22}
\delta a_{ij}^{(v,\sg)} &=& \frac{2}{3}\frac{\alpha_s}{\pi} \left\{ \right.
2\msg m_q v_q (S^{\sq})_{ij} (2C^+_{ij} + C^0_{ij})\\ 
&& + v_q \delta_{ij}\left[
(2\msg^2+ 2m_q^2+\msi^2+\msj^2)
C^+_{ij}
 +2 \msg^2 C^0_{ij} +B^0(s,m^2_q,m^2_q) \right] +a_q (A^{\sq})_{ij}
\left[\right.\nonumber\\
&&\left.(2\msg^2- 2m_q^2+\msi^2+\msj^2)
C^+_{ij}+(\msi^2-\msj^2)C^-_{ij}
 +2 \msg^2 C^0_{ij} +B^0(s,m^2_q,m^2_q) \right]  
\left.\right\}\nonumber
\, ,
\enar
and
\begin{eqnarray}
\delta {(e_q)}_{ij}^{(v, \sg)} &=& \label{eqvertex}
e_q \frac{2}{3}\frac{\alpha_s}{\pi} \left\{ \right.
 2\msg m_q (S^{\sq})_{ij}
( 2 C^+_{ij}+ C^0_{ij})\\
&&+\delta_{ij}
\left[ (2\msg^2+ 2m_q^2+\msi^2+\msj^2)
C^+_{ij}+2\msg C^0_{ij}+B^0(s,m^2_q,m^2_q) \right]
\non \left.\right\}\non \, ,
\end{eqnarray}
where $\delta_{ij}$ is the identity matrix, $v_q = 2 I_q^{3L} - 4 s_W^2 e_q$,
$a_q = 2 I_q^{3L}$,  
\bear
A^{\sq}=\left( \begin{array}{cc} \cos 2\theta_\sq & -\sin 2\theta_\sq \\
-\sin 2\theta_\sq & -\cos 2\theta_\sq
\end{array} \right) &,&
S^{\sq}=\left( \begin{array}{cc} -\sin 2\theta_\sq & -\cos 2\theta_\sq \\
-\cos 2\theta_\sq & \sin 2\theta_\sq
\end{array} \right)\, .
\enar
The functions $C^\pm_{ij}$ are defined by
\begin{equation}
C^+ = \frac{C^1+C^2}{2} \qquad , \qquad C^- = \frac{C^1-C^2}{2} \, .
\end{equation}
The arguments of all C--functions are $(m_{\sq_i}^2, s, m_{\sq_j}^2,
m_\sg^2, m_q^2, m_q^2)$. $B^0, C^0, C^1$, and $C^2$ are the usual two--
and three--point functions as given, for instance, in \cite{8}:
\bear
B^0(k^2, m_1^2, m_2^2)&=&
\int\frac{d^Dq}{i\pi^2}\frac{1}{(q^2-m_1^2)((q+k)^2-m_2^2)}\,, \nonumber\\
\left[C^0, k^\mu C^1 - \bar{k}^\mu C^2\right]&=&
\int\frac{d^Dq}{i\pi^2}\frac{[1, q^\mu]}{(q^2-m_{\sg}^2)((q+k)^2-m_q^2)
((q-\bar{k})^2-m_q^2)} \nonumber\, .
\enar
The squark wave--function renormalization due to the squark self--energy graphs
with gluino exchange (Fig.~1h) leads in the on--shell scheme to:
\bear \label{2.9}
\delta a_{ij}^{(w,\sg)}&=& \frac{1}{2} (\delta Z^{(\sg)}_{ii} + 
\delta Z_{jj}^{(\sg)}) a_{ij}
 + \delta Z^{(\sg)}_{i'i}  a_{i'j} 
+ \delta Z_{j'j}^{(\sg)}  a_{ij'} \\ 
&=&\nonumber -\mbox{Re}\left\{
\frac{1}{2}\left[\Sigma_{ii}'^{(\sg)}(m_{\sq_i}^2)
+\Sigma_{jj}'^{(\sg)}(m_{\sq_j}^2)\right] a_{ij}
+\frac{\Sigma_{i'i}^{(\sg)}(m_{\sq_i}^2)}{m_{\sq_i}^2-m_{\sq_{i'}}^2}
a_{i'j}
+\frac{\Sigma_{j'j}^{(\sg)}(m_{\sq_j}^2)}{m_{\sq_j}^2-m_{\sq_{j'}}^2}
a_{ij'}\right\}
\, ,\begin{array}{c} i \neq i' \\ j \neq j'\end{array} \, ,
\enar
and
\begin{equation} \label{eqwave}
\delta (e_q)^{(w,\sg)}_{ii} = -e_q \mbox{Re}\left\{
\Sigma_{ii}'^{(\sg)}(m_{\sq_i}^2) \right\}\,, \quad 
\delta (e_q)^{(w,\sg)}_{12} = \frac{e_q}{m_{\sq_1}^2-m_{\sq_{2}}^2}
\mbox{Re}\left\{ \Sigma_{12}^{(\sg)}(m_{\sq_2}^2) -
\Sigma_{21}^{(\sg)}(m_{\sq_1}^2) \right\} \, ,
\eneq
with the squark self--energy contributions 
$\Sigma_{ij}^{(\sg)}(m^2)$ and their derivatives\\
$\Sigma_{ii}'^{(\sg)}(m^2)= \partial \Sigma_{ii}^{(\sg)}(p^2)/
\partial p^2 |_{p^2=m^2}$:
\beeq
\Sigma_{12}^{(\sg)}(p^2) = \Sigma_{21}^{(\tilde g)}(p^2) =
\frac{4}{3}\frac{\alpha_s}{\pi} \msg m_q \cos 2 \heta 
B_0(p^2, \msg^2,m_q^2)
\eneq
and
\begin{eqnarray} \label{29}
\Sigma_{ii}^{'\,(\tilde g)} (p^2) &=&
\frac{2}{3}\frac{\alpha_s}{\pi} \left[
B_0(p^2, \msg^2,m_q^2) + (p^2-m_q^2-\msg^2)
B_0'(p^2, \msg^2,m_q^2) \right. \non \\
& & \left. \hspace*{1cm} -2 m_q \msg (-1)^i \sin 2 \heta
B_0'(p^2, \msg^2,m_q^2) \right] \, .
\end{eqnarray}

\subsection{The renormalization of the squark mixing angle}
We still have to discuss the renormalization of the squark mixing angle
$\delta a_{ij}^{(\tilde\theta,\sq)}$, eq.~(\ref{deltaasq}), and
$\delta a_{ij}^{(\tilde\theta,\sg)}$, eq.~(\ref{deltaasg}).
From eq.~(\ref{acoup}) one gets for $\delta a_{ij}^{(\tilde\theta,a)}$ (with
$a = \sq, \sg$):
\beeq \label{27}
\delta a_{11}^{(\tilde\theta, a)} = 2 a_{12} \delta \heta^{(a)} =
-\delta a_{22}^{(\tilde\theta, a)}\quad \, \quad
\delta a_{12}^{(\tilde\theta, a)} = \delta a_{21}^{(\tilde\theta, a)}
= (a_{22} - a_{11}) \delta \heta^{(a)}\, .
\eneq
Now $\Delta a_{ij}^{(\sq)}$ of eq.~(\ref{deltaasq})
has to be free from ultra--violet divergencies and is therefore
finite. Choosing $\Delta a_{12}^{(\sq)} = 0$ one gets from eqs.~(\ref{deltaasq})
and (\ref{15}) 
\beeq \label{da12thetaq}
\delta a_{12}^{(\tilde\theta, \tilde q)} =
(a_{22} - a_{11}) \delta\heta^{(\tilde q)} =
- \delta Z_{21}^{(\tilde q)} a_{22} -
\delta Z_{12}^{(\tilde q)} a_{11}\, .
\eneq  
Eq.~(\ref{da12thetaq}) means that the related counterterm 
$\delta a_{12}^{(\tilde\theta,\sq)}$ is nothing else than the negative
sum of the graphs corresponding to Fig.~1j containing the self--energies
$\Sigma_{12}^{(\sq)}$ and $\Sigma_{21}^{(\sq)}$. We now do the same for
$\delta a_{ij}^{(\tilde\theta,\sg)}$ by taking again the negative sum
of those parts of the graphs Fig.~1h, which contain the self--energies
$\Sigma_{12}^{(\sg)}(m_{\sq_2}^2)$ and $\Sigma_{21}^{(\sg)}(m_{\sq_1}^2)$.
$\delta a_{12}^{(\tilde\theta,\sg)}$ is then also given by
\beeq \label{da12thetag}
\delta a_{12}^{(\tilde\theta, \tilde g)} =
(a_{22} - a_{11}) \delta\heta^{(\tilde g)} =
- \delta Z_{21}^{(\tilde g)} a_{22} -
\delta Z_{12}^{(\tilde g)} a_{11}\, ,
\eneq 
with
\begin{equation}
\delta Z_{12}^{(\sg)}=
\frac{\mbox{Re}\left\{\Sigma_{12}^{(\sg)}(m_{\tilde q_2}^2)\right\}
}{m_{\tilde q_1}^2-
m_{\tilde q_2}^2} \quad , \quad
\delta Z_{21}^{(\sg)}=
\frac{\mbox{Re}\left\{\Sigma_{21}^{(\sg)}(m_{\tilde q_1}^2)\right\}
}{m_{\tilde q_2}^2-
m_{\tilde q_1}^2} \, .
\end{equation}
Inserting the results for $\delta Z_{12}^{(\tilde q)},
\delta Z_{21}^{(\tilde q)}$ in eq.~(\ref{da12thetaq}) and
$\delta Z_{12}^{(\tilde g)}, \delta Z_{21}^{(\tilde g)}$ in 
eq.~(\ref{da12thetag}), one gets
\beeq \label{dthsq}
\delta\heta^{(\sq)}
= \frac 1 6 \frac{\alpha_s}{\pi} \frac{\sin 4 \heta}{m_{\sq_1}^2 -
m_{\sq_2}^2}\left( A^0(m_{\sq_2}^2) - A^0(m_{\sq_1}^2)\right)\, ,
\eneq
and
\beeq \label{dtheta}
\delta\theta_{\tilde q}^{(\tilde g)} =
\frac{1}{3}\frac{\alpha_s}{\pi}
\frac{m_{\sg} m_q}{I^{3L}_q (m_{\sq_1}^2 - m_{\sq_2}^2)}\left(
B^0(m_{\sq_2}^2,m_{\sg}^2,m_q^2) a_{11} -
B^0(m_{\sq_1}^2,m_{\sg}^2,m_q^2) a_{22}\right) \, .
\eneq
By eqs.~(\ref{27}) we then obtain $\delta a_{ij}^{(\tilde\theta,\sq)}$ and
$\delta a_{ij}^{(\tilde\theta,\sg)}$. 
With this result for $\delta a_{ij}^{(\tilde\theta,\sq)}$ there is no
correction from the four--squark interaction graphs: $\Delta a_{ij}^{(\sq)}
= 0$, and due to $\Delta (e_q)_{ij}^{(\sq)} = 0$ also $\delta\sigma^\sq = 0$.\\
By knowing $\delta a_{ij}^{(\tilde\theta,\sg)}$ we can write the final result
for the total gluino correction:
\begin{eqnarray} \label{32}
\Delta a_{ij}^{(\sg)} &=& \delta a_{ij}^{(v,\sg)} -\mbox{Re}\left\{\right.
\frac{1}{2}
\left(\Sigma_{ii}'(m_{\sq_i}^2)
+\Sigma_{jj}'(m_{\sq_j}^2)\right) a_{ij}
+ \frac{4}{3}\frac{\alpha_s}{\pi}
\frac{m_{\sg} m_q}{m_{\sq_1} - m_{\sq_2}}\delta_{ij}\\
&\times &\left[
B^0(m_{\sq_i}^2,m_{\sg}^2,m_q^2)((-1)^{i+1} 2 a_{ii'} \cos 2\theta_\sq -
a_{i'i'} \sin 2\theta_\sq) +
B^0(m_{\sq_{i'}}^2,m_{\sg}^2,m_q^2) a_{ii} \sin 2\theta_\sq\right]
\left.\right\}\non\, ,
\end{eqnarray}
with $i' \neq i$, and $\Delta {(e_q)}_{ij}^{(\sg)} =
\delta {(e_q)}_{ij}^{(v,\sg)} +\delta {(e_q)}_{ij}^{(w,\sg)}$ as in
eq.~(\ref{deltaesg}),
where $\delta a_{ij}^{(v,\sg)}, \Sigma_{ii}'(m_{\sq_i}^2),
\delta {(e_q)}_{ij}^{(v,\sg)}$, and
$\delta {(e_q)}_{ij}^{(w,\sg)}$ are given by
eqs.~(\ref{22}), (\ref{29}), (\ref{eqvertex}), and (\ref{eqwave}), respectively.
Note that $\Delta a_{ij}^{(\sg)}$ and $\Delta {(e_q)}_{ij}^{(\sg)}$ are
ultra--violet finite.\\
Inserting $\Delta a_{ij}^{(\sg)}$ and $\Delta (e_q)_{ij}^{(\sg)}$ in
eq.~(\ref{dsiga})
gives the total gluino correction $\delta\sigma^\sg$ which
does not factorize as the gluon correction in eq.~(\ref{dsigg}).
  
Our renormalization condition for the mixing angle $\heta$ is different from
that of \cite{7}. For $\heta = 0$ and $m_{\sq_i} = m_{\sq_j}$ our results agree
with those of \cite{7}. However, in the scheme of \cite{7} there appears a
singularity at $\heta = 45^\circ$ going with 
 $\sim$ $\tan 2\theta_{\tilde q}$ in the diagonal channels $e^+ e^- \to \sq_1 
\bar{\sq}_1,\,\sq_2 \bar{\sq}_2$, so that
this renormalization scheme is numerically not applicable in the region around
$\heta = 45^\circ$.
In our scheme there is no such singularity as can be seen in
eq.~(\ref{dtheta}) so that it can be applied in the whole
range of $\heta$.
 
\section{Numerical results}

We now turn to the numerical analysis of the SUSY--QCD corrections. For the
gluonic correction $\delta\sigma^g$ we evaluate eqs.~(\ref{dsigg}),
(\ref{delgluon}) and (\ref{lam}), and for the
correction due to gluino exchange $\delta\sigma^\sg$ we take eqs.~(\ref{32})
and (\ref{deltaesg}) together with eq.~(\ref{dsiga}). In the following
we have always taken
$m_t = 175$~GeV.\\

We have first calculated the corrections to the cross section 
$\sigma(e^+ e^- \to \st_1 \bar\st_1)$ at the LEP2 energy $\sqrt{s} = 190$~GeV
as a function of the stop mass $m_{\st_1}$ for $\cos\theta_\st = 0.7$, 
taking $m_{\sg} = 200$~GeV and $m_{\st_2} = 250$~GeV.
This is shown in
Fig.~2. The gluon correction is rising from 17\% of $\sigma^{tree}$ for
$m_{\st_1} = 45$~GeV up to 35\% for $m_{\st_1} = 80$~GeV. The gluino
correction is only about 1\% and quite independent of $m_{\st_1}$.\\

In Fig.~3a we show the corrections to the cross section $\sigma(e^+ e^-
\to \st_1 \bar\st_1)$
as a function of the mixing angle $\cos\theta_\st$, for $\sqrt{s} = 500$~GeV,
$m_{\st_1} = 150$~GeV, $m_{\st_2} = 300$~GeV, and $m_{\sg} = 300$~GeV.
According to eqs.~(\ref{dsigg}) and (\ref{delgluon}) 
$\delta\sigma^g$ has the same
dependence on $\cos\theta_\st$ as the tree--level cross section, whereas the
gluino correction (see eqs.~(\ref{32})) introduces a different 
$\theta_\st$ dependence. This is due to the fact that the gluino couples
equally to $\st_L^* t_L$ and to $\st_R^* t_R$ and therefore, in the case of 
mixing, the couplings $\sg \st_i^* t$ 
get a dependence on the mixing angle $\theta_\st$.
For the 
c. m. energy and masses chosen the gluino correction is
now higher than at LEP2 energies and is about 15\% of
the gluon correction.
Fig.~3b exhibits the $\sqrt{s}$ dependence of $\delta\sigma^g/\sigma^{tree}$
and $\delta\sigma^\sg/\sigma^{tree}$ for $\cos\theta_\st = 0.7$ and 
keeping the masses as in Fig.~3a. Notice that $\delta\sigma^g/\sigma^{tree}$
is decreasing with $\sqrt{s}$, whereas $\delta\sigma^\sg/\sigma^{tree}$ is
becoming negative at $\sqrt{s} = 700$~GeV with the absolute value increasing
with $\sqrt{s}$.\\
Moreover, we have examined the dependences of 
$\delta\sigma^{\sg}/\sigma^{tree}$ in 
$e^+ e^- \to \st_1 \bar\st_1$ at fixed $\sqrt{s}$ on the squark masses
$m_{\st_1}$, $m_{\st_2}$, and on their difference. These dependences are weak.\\ 

Fig.~4a and b show the $\cos\theta_\st$ dependence 
for a higher energy and mass scenario,
$\sqrt{s} = 2$~TeV, 
$m_{\st_1} = 500$~GeV, $m_{\st_2} = 700$~GeV, and $m_{\sg} = 600$~GeV.
Here $\delta\sigma^\sg$ is about $-35$\% 
of $\delta\sigma^g$ for $\st_1 \bar\st_1$
production.
Fig.~4b exhibits the $\st_1 \bar\st_2$ channel.
$\delta\sigma^g/\sigma^{tree}$ is about 16\%. 
$\delta\sigma^\sg$ is about $-40$\% of $\delta\sigma^g$ at the peak values
Notice that 
the gluino correction is not always positive. In Fig.~4c we
also show for this scenario the corrections in the case of the $\st_2 \bar\st_2$
production. Here the gluino part is even larger ($-50$\% of the gluon 
contribution). \\

We have also computed the SUSY--QCD corrections to $\sigma(e^+ e^- \to
\sb_1 \bar\sb_1)$. For $\sqrt{s} = 2$~TeV, 
$m_{\sb_1} = 500$~GeV, $m_{\sb_2} = 700$~GeV, and $m_{\sg} = 600$~GeV 
$\delta\sigma^\sg$ is about
$-30$\% of $\delta\sigma^g$ almost independent of $\cos\theta_\sb$.
The gluon correction for $\sb_i \bar\sb_j$ 
must be the same
as that for $\st_i \bar\st_j$ production,
provided the squark masses are the same. 
This can also be seen explicitely in
eq.~(\ref{delgluon}).\\

It is particularly interesting to study the dependence on the gluino mass. This
can be seen in Fig.~5, where we have 
$\sqrt{s} = 500$~GeV, $m_{\st_1} = 150$~GeV,
$m_{\st_2} = 300$~GeV, and $\cos\theta_\st = 0.5$ in Fig.~5a,
and $\sqrt{s} = 2$~TeV, $m_{\st_1} = 500$~GeV, $m_{\st_2} = 700$~GeV, and
$\cos\theta_\st = 0.5$ in Fig.~5b. It is somewhat surprising that the gluino
correction is first increasing as a function of the gluino mass and only very
slowly decreasing. Of course the correction would be largest for a small gluino
mass ($m_\sg \lsim 50$~GeV) but this is excluded by experiment 
($m_\sg \gsim 130$~GeV).\\
Here we also want to notice that threshold singularities appear when $m_{\st_i}
= m_t + m_\sg$. For instance, such a singularity could appear in Fig.~5b for
$m_\sg = 325$~GeV. These singularities stem from the self--energy functions
in the wave--function renormalization of the stops. This is a known effect and
has been also observed in other cases \cite{jimenez}.\\

In conclusion, our analysis of the SUSY--QCD corrections to scalar
quark pair production in $e^+ e^-$ annihilation has shown that the
correction due to the gluino exchange are smaller than the conventional
QCD corrections, but they are 
significant at energies envisaged for the next
linear collider \cite{NLC}. In particular, they exhibit a
dependence on the mixing angle $\theta_\sq$, which is different from the 
tree--level cross section corrected by the gluon exchange. Moreover, the
correction due to gluino exchange decreases only very slowly with increasing 
gluino mass.

\vspace{5mm}
\section*{Acknowledgements}
This work grew out of the Workshop on Physics at LEP2, CERN, 1995, and
the Workshop on Physics with $e^+ e^-$ Linear Colliders, Annecy -- Gran Sasso --
Hamburg, 1995. 
We thank the members of
the ``Supersymmetry Working Groups'' for discussions. This 
work was supported by the ``Fonds zur F\"orderung der wissenschaftlichen
Forschung'' of Austria, project no. P10843--PHY.

%\newpage

\newpage
%\vspace{20mm}
\section*{Figure Captions}
\renewcommand{\labelenumi}{Fig.~\arabic{enumi}} \begin{enumerate}

\vspace{6mm}
\item
Feynman diagrams for the lowest order SUSY--QCD corrections to
$e^+ e^- \to \sq_i \bar{\sq}_j$. Note that there are also the corresponding
diagrams to c), d), e), h), and j) for the antisquark $\bar\sq_j$.

\vspace{6mm}
\item
SUSY--QCD corrections $\delta\sigma^g/\sigma^{tree}$ and 
$\delta\sigma^\sg/\sigma^{tree}$
as a function of $m_{\st_1}$ for
$e^+ e^- \to \st_1 \bar\st_1$
for $\sqrt{s} = 190$~GeV, $\cos\theta_\st = 0.7$, $m_{\st_2} = 250$~GeV, and
$m_{\sg} = 200$~GeV.

\vspace{6mm}
\item
SUSY--QCD corrections $\delta\sigma^g/\sigma^{tree}$ and 
$\delta\sigma^\sg/\sigma^{tree}$ 
for $e^+ e^- \to \st_1 \bar\st_1$\\
(a) as a function of $\cos\theta_\st$ for
$\sqrt{s} = 500$~GeV, $m_{\st_1} = 150$~GeV, $m_{\st_2} = 300$~GeV, and
$m_{\sg} = 300$~GeV,\\
(b) as a function of $\sqrt{s}$ for $\cos\theta_\st = 0.7$, 
$m_{\st_1} = 150$~GeV, $m_{\st_2} = 300$~GeV, and
$m_{\sg} = 300$~GeV. 

\vspace{6mm}
\item
SUSY--QCD corrections $\delta\sigma^g$ and $\delta\sigma^\sg$ as a function
of $\cos\theta_\st$ for\\
$\sqrt{s} = 2$~TeV, $m_{\st_1} = 500$~GeV, $m_{\st_2} = 700$~GeV, and
$m_{\sg} = 600$~GeV.\\
(a) for $e^+ e^- \to \st_1 \bar\st_1$\\
(b) for $e^+ e^- \to \st_1 \bar\st_2$\\
(c) for $e^+ e^- \to \st_2 \bar\st_2$

\vspace{6mm}
\item
Dependence of the SUSY--QCD corrections $\delta\sigma^g/\sigma^{tree}$ and 
$\delta\sigma^{g+\sg}/\sigma^{tree}$ 
on the gluino mass for $e^+ e^- \to \st_1 \bar\st_1$.\\
(a) for $\sqrt{s} = 500$~GeV, $m_{\st_1} = 150$~GeV, $m_{\st_2} = 300$~GeV,
$\cos\theta_\st = 0.5$\\  
(b) for $\sqrt{s} = 2$~TeV, $m_{\st_1} = 500$~GeV, $m_{\st_2} = 700$~GeV,
$\cos\theta_\st = 0.5$  

\end{enumerate}

\clearpage
%Figure 1: PAGE with Feynman graphs:
%\clearpage
\setlength{\unitlength}{1mm}

\begin{center}
\begin{picture}(170,210)(-7,0)
%\put(0,0){\framebox(170,210){}}
%\put(0,0){\graphpaper[2](0,0)(170,210)}
\put(0,15){\mbox{\epsfig{file=e+e-sq.ps,height=19.5cm}}}
%figure 1a:
\put(78.5,210){\makebox(0,0)[t]{\large{\bf{a)}}}}
\put(53,210){\makebox(0,0)[r]{$e^-$}}
\put(53,179){\makebox(0,0)[r]{$e^+$}}
\put(78.5,197){\makebox(0,0)[b]{$\gamma, Z^0$}}
\put(104,210){\makebox(0,0)[l]{$\sq_i$}}
\put(104,179){\makebox(0,0)[l]{$\bar{\sq}_j$}}
%figure 1b:
\put(1.5,169){\makebox(0,0)[tl]{\large{\bf{b)}}}}
\put(1.5,155.5){\makebox(0,0)[bl]{$\gamma, Z^0$}}
\put(39,169){\makebox(0,0)[l]{$\sq_i$}}
\put(32.5,153.5){\makebox(0,0)[l]{$g$}}
\put(39,138){\makebox(0,0)[l]{$\bar{\sq}_j$}}
%figure 1c:
\put(60,169){\makebox(0,0)[tl]{\large{\bf{c)}}}}
\put(60,155.5){\makebox(0,0)[bl]{$\gamma, Z^0$}}
\put(98,169){\makebox(0,0)[l]{$\sq_i$}}
\put(79,162){\makebox(0,0)[br]{$g$}}
\put(98,138){\makebox(0,0)[l]{$\bar{\sq}_j$}}
%figure 1d:
\put(119,169){\makebox(0,0)[tl]{\large{\bf{d)}}}}
\put(119.5,155.5){\makebox(0,0)[bl]{$\gamma, Z^0$}}
\put(157,169){\makebox(0,0)[l]{$\sq_i$}}
\put(142.5,166){\makebox(0,0)[br]{$g$}}
\put(157,138){\makebox(0,0)[l]{$\bar{\sq}_j$}}
%figure 1e:
\put(44,128){\makebox(0,0)[t]{\large{\bf{e)}}}}
\put(18.5,128){\makebox(0,0)[r]{$e^-$}}
\put(18.5,97){\makebox(0,0)[r]{$e^+$}}
\put(44,114.5){\makebox(0,0)[b]{$\gamma, Z^0$}}
\put(69,128){\makebox(0,0)[l]{$\sq_i$}}
\put(69,120){\makebox(0,0)[l]{$g$}}
\put(69,97){\makebox(0,0)[l]{$\bar{\sq}_j$}}
%figure 1f:
\put(114,128){\makebox(0,0)[t]{\large{\bf{f)}}}}
\put(87.5,128){\makebox(0,0)[r]{$e^-$}}
\put(87.5,97){\makebox(0,0)[r]{$e^+$}}
\put(114,114.5){\makebox(0,0)[b]{$\gamma, Z^0$}}
\put(138.5,128){\makebox(0,0)[l]{$\sq_i$}}
\put(138.5,112.5){\makebox(0,0)[l]{$g$}}
\put(138.5,97){\makebox(0,0)[l]{$\bar{\sq}_j$}}
%figure 1g:
\put(30.5,86.5){\makebox(0,0)[tl]{\large{\bf{g)}}}}
\put(30.5,73.5){\makebox(0,0)[bl]{$\gamma, Z^0$}}
\put(55,76){\makebox(0,0)[br]{$q$}}
\put(61.5,71.5){\makebox(0,0)[l]{$\sg$}}
\put(55,67){\makebox(0,0)[tr]{$\bar{q}$}}
\put(68,86.5){\makebox(0,0)[l]{$\sq_i$}}
\put(68,56.5){\makebox(0,0)[l]{$\bar{\sq}_j$}}
%figure 1h:
\put(90,86.5){\makebox(0,0)[tl]{\large{\bf{h)}}}}
\put(90,73.5){\makebox(0,0)[bl]{$\gamma, Z^0$}}
\put(119,78){\makebox(0,0)[tl]{$q$}}
\put(114,84){\makebox(0,0)[br]{$\sg$}}
\put(112,75){\makebox(0,0)[br]{$\sq_k$}}
\put(127.5,86.5){\makebox(0,0)[l]{$\sq_i$}}
\put(127.5,56.5){\makebox(0,0)[l]{$\bar{\sq}_j$}}
%figure 1i:
\put(30.5,46){\makebox(0,0)[tl]{\large{\bf{i)}}}}
\put(30.5,32.5){\makebox(0,0)[bl]{$\gamma, Z^0$}}
\put(47,37.5){\makebox(0,0)[b]{$\sq_{1,2}$}}
\put(47,24){\makebox(0,0)[t]{$\bar{\sq}_{1,2}$}}
\put(68,46){\makebox(0,0)[l]{$\sq_i$}}
\put(68,15){\makebox(0,0)[l]{$\bar{\sq}_j$}}
%figure 1j:
\put(90,46){\makebox(0,0)[tl]{\large{\bf{j)}}}}
\put(90,32.5){\makebox(0,0)[bl]{$\gamma, Z^0$}}
\put(115,49){\makebox(0,0)[br]{$\sq_{1,2}$}}
\put(115,33){\makebox(0,0)[tl]{$\sq_k$}}
\put(127.5,46){\makebox(0,0)[l]{$\sq_i$}}
\put(127.5,15){\makebox(0,0)[l]{$\bar{\sq}_j$}}
\put(78.5,0){\makebox(0,0)[b]{\Large{\bf{Fig.~1}}}}
\end{picture}

\large

\newpage
%figures 2:
\begin{picture}(85,85)(0,0)
%\put(0,0){\framebox(85,85){}}
\put(46,-8){\makebox(0,0)[t]{\Large{\bf{Fig.~2}}}}
%\graphpaper[5](0,0)(85,85)
\put(0,0){\mbox{\epsfig{file=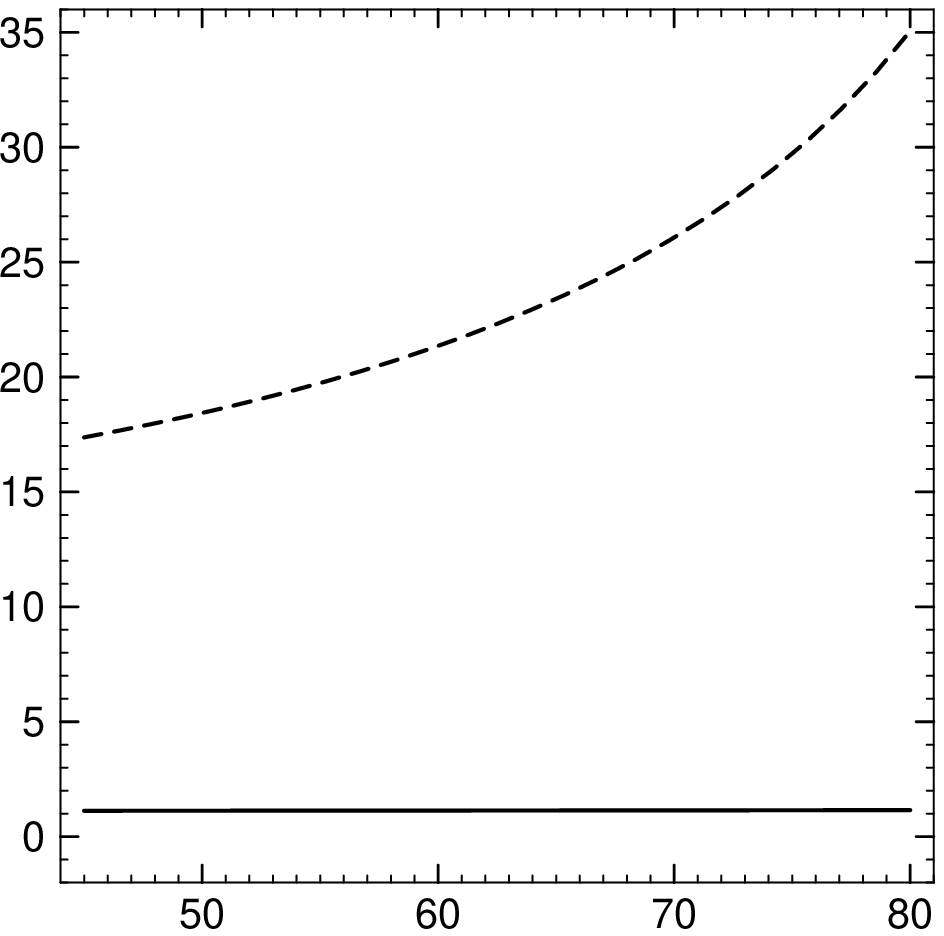,height=85mm}}}
\put(47,-1){\makebox(0,0)[t]{{\boldmath\large{$m_{\st_1}$}}}}
\put(-2,82){\makebox(0,0)[tr]{\bf{\%}}}
\put(47,35){\makebox(0,0)[b]{\boldmath{$e^+ e^- \to \tilde{t}_1
\bar{\tilde{t}}_1$}}}
\put(47,59){\makebox(0,0)[b]{$\delta\sigma^g/\sigma^{tree}$}}
\put(47,12){\makebox(0,0)[b]{$\delta\sigma^{\tilde g}/
\sigma^{tree}$}}
\end{picture}

\vfill

%figures 3a:
\begin{picture}(85,85)(0,0)
%\put(0,0){\framebox(85,85){}}
\put(46,-8){\makebox(0,0)[t]{\Large{\bf{Fig.~3a}}}}
%\graphpaper[5](0,0)(85,85)
\put(0,0){\mbox{\epsfig{file=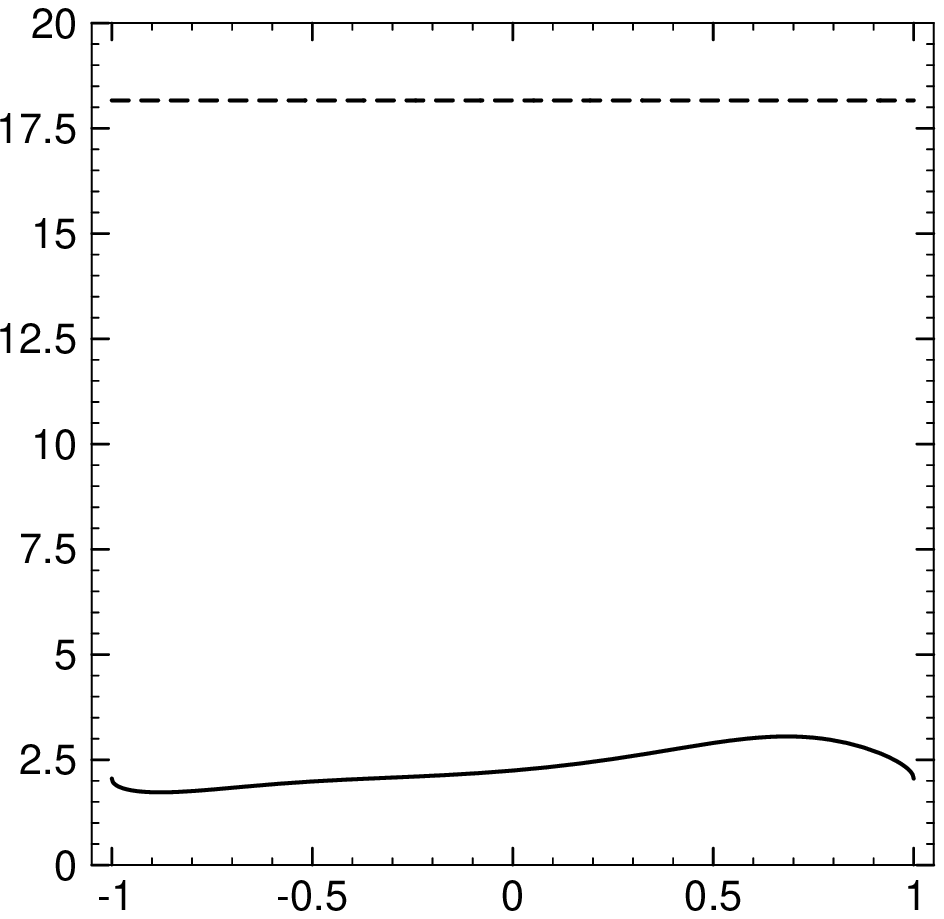,height=85mm}}}
\put(47,0){\makebox(0,0)[t]{{\boldmath\large{$\cos\theta_{\tilde{t}}$}}}}
\put(-2,82){\makebox(0,0)[tr]{\bf{\%}}}
\put(47,50){\makebox(0,0)[b]{\boldmath{$e^+ e^- \to \tilde{t}_1
\bar{\tilde{t}}_1$}}}
\put(47,73){\makebox(0,0)[t]{$\delta\sigma^g/\sigma^{tree}$}}
\put(47,20){\makebox(0,0)[t]{$\delta\sigma^{\tilde g}/
\sigma^{tree}$}}
\end{picture}

\vspace{14mm}

\newpage
%figures 3b:
\begin{picture}(85,85)(0,0)
%\put(0,0){\framebox(85,85){}}
%\graphpaper[5](0,0)(85,85)
\put(46,-8){\makebox(0,0)[t]{\Large{\bf{Fig.~3b}}}}
\put(0,0){\mbox{\epsfig{file=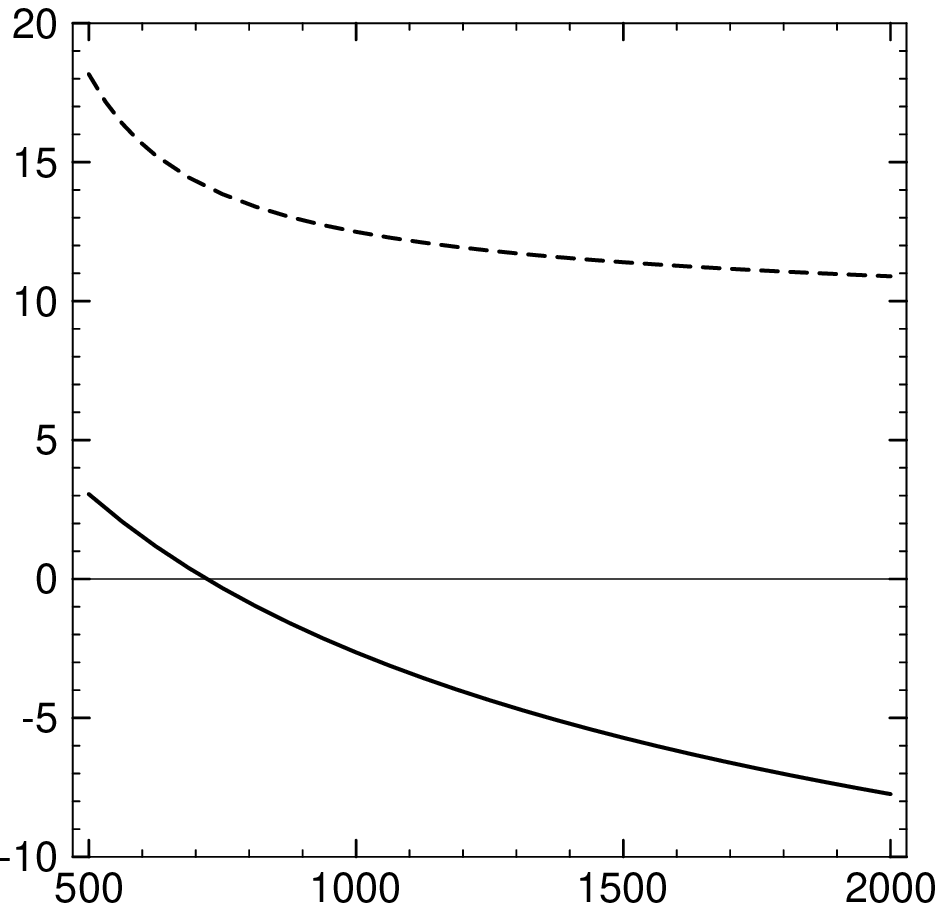,height=85mm}}}
\put(46,0){\makebox(0,0)[t]{{\boldmath\large{$\sqrt{s}$}}~\normalsize{[GeV]}}}
\put(-2,83){\makebox(0,0)[tr]{\bf{\%}}}
\put(46,45){\makebox(0,0)[b]{\boldmath{$e^+ e^- \to \tilde{t}_1
\bar{\tilde{t}}_1$}}}
\put(17,68){\makebox(0,0)[bl]{$\delta\sigma^g/\sigma^{tree}$}}
\put(66.5,16){\makebox(0,0)[b]{$\delta\sigma^{\tilde g}/\sigma^{tree}$}}
\end{picture}

\vfill

%figures 4a:
\begin{picture}(85,85)(0,0)
%\put(0,0){\framebox(85,85){}}
%\graphpaper[5](0,0)(85,85)
\put(46,-8){\makebox(0,0)[t]{\Large{\bf{Fig.~4a}}}}
\put(0,0){\mbox{\epsfig{file=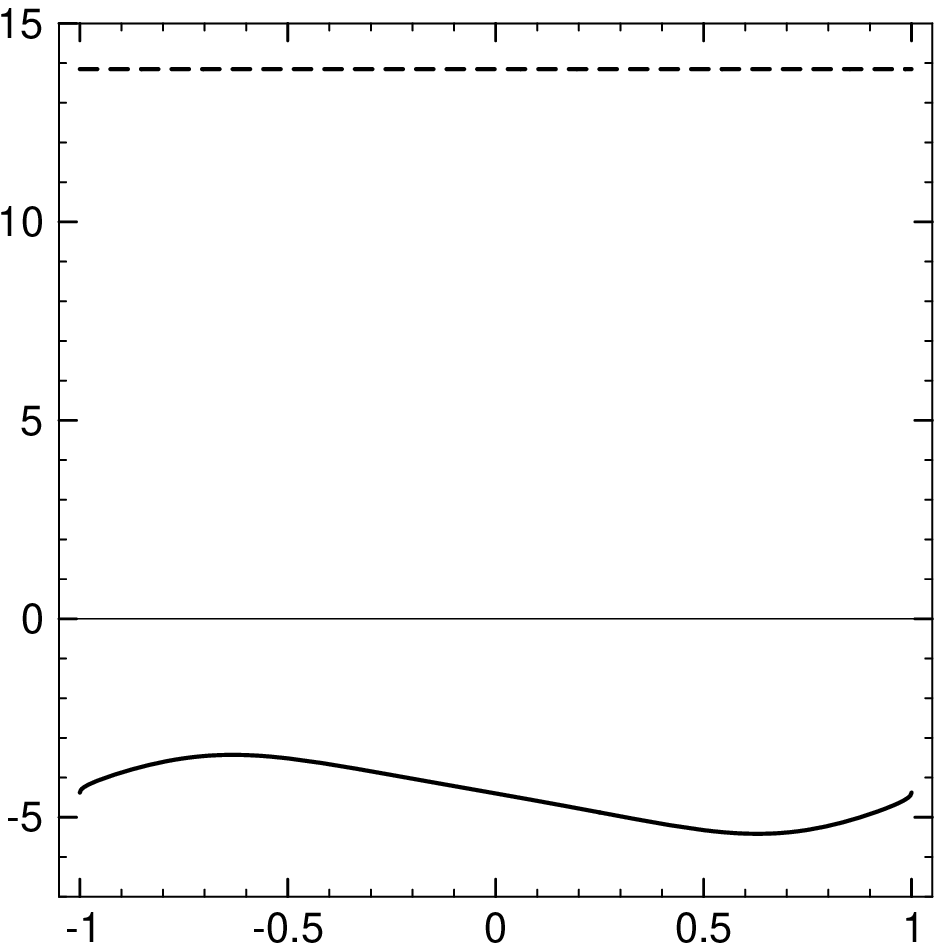,height=85mm}}}
\put(45,-1){\makebox(0,0)[t]{{\boldmath\large{$\cos\theta_{\tilde{t}}$}}}}
\put(-3,84){\makebox(0,0)[tr]{\bf{\%}}}
\put(45,50){\makebox(0,0)[b]{\boldmath{$e^+ e^- \to \tilde{t}_1
\bar{\tilde{t}}_1$}}}
\put(45,77){\makebox(0,0)[t]{$\delta\sigma^g/\sigma^{tree}$}}
\put(42,15){\makebox(0,0)[bl]{$\delta\sigma^{\tilde g}/
\sigma^{tree}$}}
\end{picture}

\vspace{14mm}

\newpage

%figures 4b:
\begin{picture}(85,85)(0,0)
%\put(0,0){\framebox(85,85){}}
%\graphpaper[5](0,0)(85,85)
\put(46,-8){\makebox(0,0)[t]{\Large{\bf{Fig.~4b}}}}
\put(0,0){\mbox{\epsfig{file=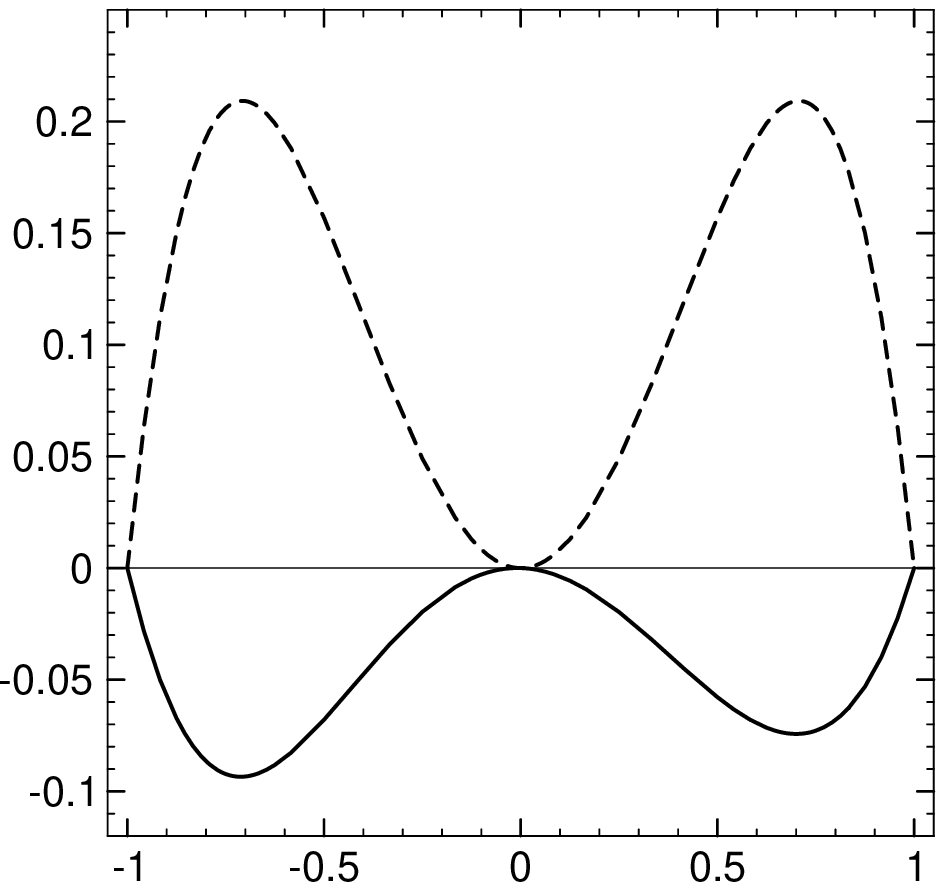,height=85mm}}}
\put(47,0){\makebox(0,0)[t]{{\boldmath\large{$\cos\theta_{\tilde{t}}$}}}}
\put(0,82){\makebox(0,0)[tr]{[fb]}}
\put(47,70){\makebox(0,0)[b]{\boldmath{$e^+ e^- \to \tilde{t}_1
\bar{\tilde{t}}_2$}}}
\put(32,60){\makebox(0,0)[bl]{$\delta\sigma^g$}}
\put(30,18){\makebox(0,0)[tl]{$\delta\sigma^{\tilde g}$}}
\end{picture}

\vfill

%figures 4c:
\begin{picture}(85,85)(0,0)
%\put(0,0){\framebox(85,85){}}
%\graphpaper[5](0,0)(85,85)
\put(46,-8){\makebox(0,0)[t]{\Large{\bf{Fig.~4c}}}}
\put(0,0){\mbox{\epsfig{file=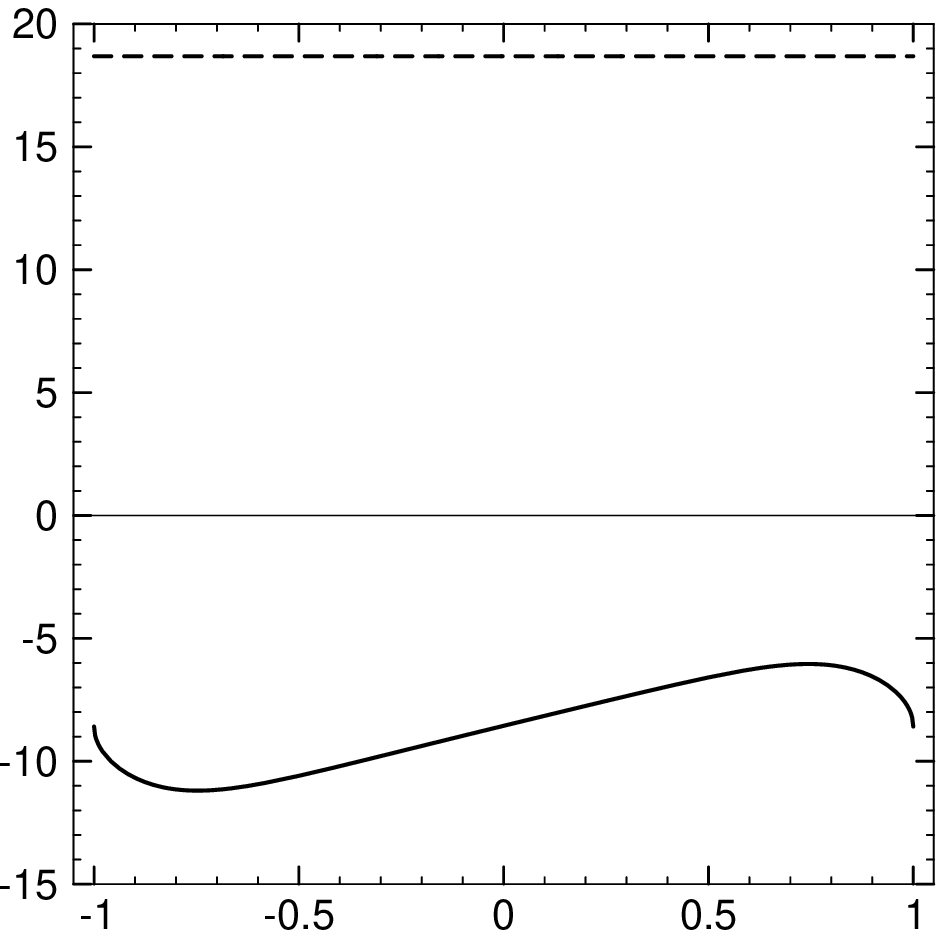,height=85mm}}}
\put(46,-1){\makebox(0,0)[t]{{\boldmath\large{$\cos\theta_{\tilde{t}}$}}}}
\put(-3,83){\makebox(0,0)[tr]{\bf{\%}}}
\put(46,50){\makebox(0,0)[b]{\boldmath{$e^+ e^- \to \tilde{t}_2
\bar{\tilde{t}}_2$}}}
\put(46,77.5){\makebox(0,0)[t]{$\delta\sigma^g/\sigma^{tree}$}}
\put(40,17){\makebox(0,0)[br]{$\delta\sigma^{\tilde g}/
\sigma^{tree}$}}
\end{picture}

\vspace{14mm}

\newpage

%figures 5a:
\begin{picture}(85,85)(0,0)
%\put(0,0){\framebox(85,85){}}
%\graphpaper[5](0,0)(85,85)
\put(0,0){\mbox{\epsfig{file=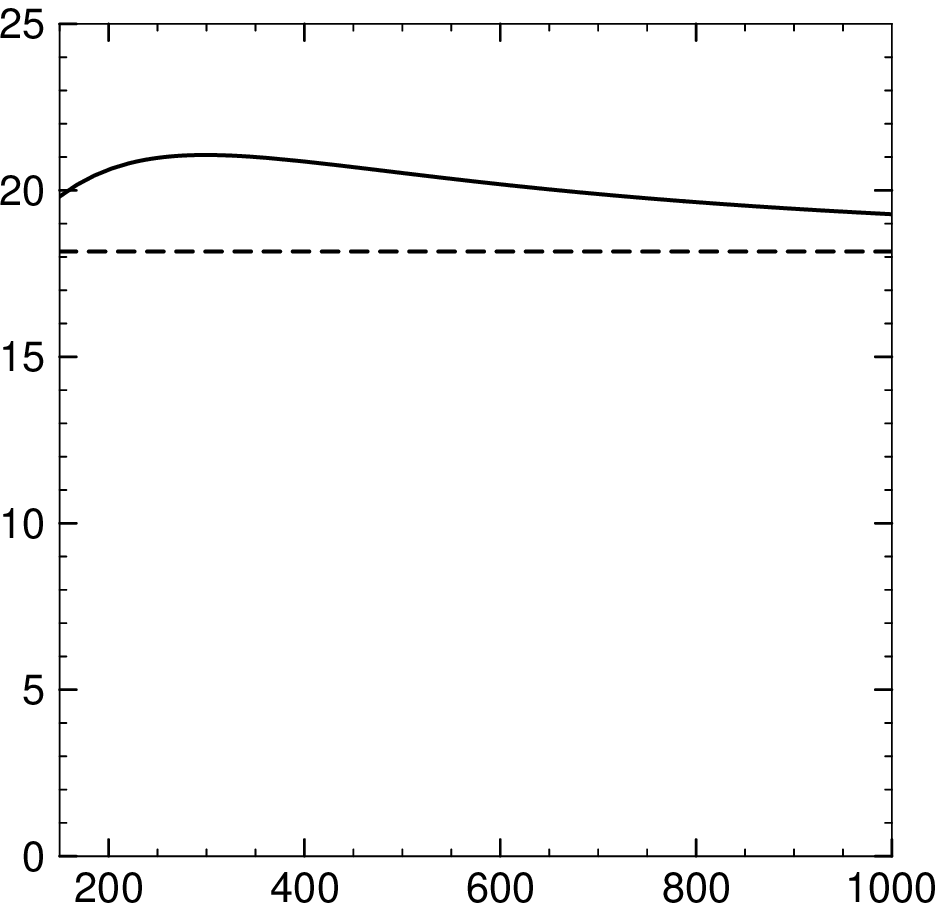,height=85mm}}}
\put(43,-8){\makebox(0,0)[t]{\Large{\bf{Fig.~5a}}}}
\put(43,0){\makebox(0,0)[t]{{\boldmath\large{$m_{\tilde{g}}$}}~
\normalsize{[GeV]}}}
\put(-4,82){\makebox(0,0)[tr]{\bf{\%}}}
\put(43,30){\makebox(0,0)[b]{\boldmath{$e^+ e^- \to \tilde{t}_1
\bar{\tilde{t}}_1$}}}
\put(45,69){\makebox(0,0)[bl]{$\delta\sigma^{g
+ \tilde g}/\sigma^{tree}$}}
\put(43,59){\makebox(0,0)[t]{$\delta\sigma^g/
\sigma^{tree}$}}
\end{picture}

\vfill

%figures 5b:
\begin{picture}(85,85)(0,0)
%\put(0,0){\framebox(85,85){}}
%\graphpaper[5](0,0)(85,85)
\put(43,-8){\makebox(0,0)[t]{\Large{\bf{Fig.~5b}}}}
\put(0,0){\mbox{\epsfig{file=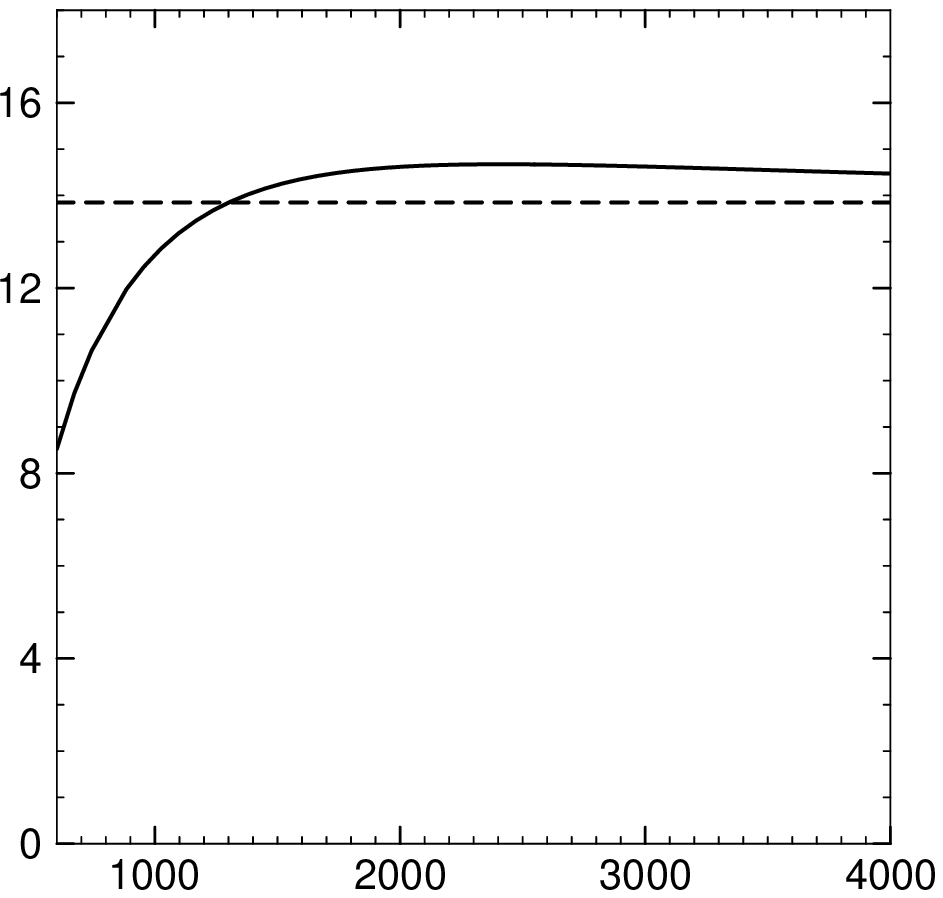,height=85mm}}}
\put(43,1){\makebox(0,0)[t]{{\boldmath\large{$m_{\tilde{g}}$}}~
\normalsize{[GeV]}}}
\put(-4,82.5){\makebox(0,0)[tr]{\bf{\%}}}
\put(43,30){\makebox(0,0)[b]{\boldmath{$e^+ e^- \to \tilde{t}_1
\bar{\tilde{t}}_1$}}}
\put(63,63){\makebox(0,0)[t]{$\delta\sigma^g/\sigma^{tree}$}}
\put(43,70){\makebox(0,0)[b]{$\delta\sigma^{g+\tilde g}/
\sigma^{tree}$}}
\end{picture}
\end{center}
\vspace{10mm}

\end{document}